\newcommand{\LASCUO}{{$\mathrm{La_{1.85}Sr_{0.15}CuO_{4+\delta}}$}}
\newcommand{\LACUO}{{$\mathrm{La_{2}CuO_{4+\delta}}$}}
\newcommand{\LASNIO}{{$\mathrm{La_{2-x}Sr_{x}NiO_{4}}$}}
\newcommand{\LASMNO}{{$\mathrm{La_{1-x}Sr_{x}MnO_{3}}$}}
\newcommand{\BKBIO}{{$\mathrm{Ba_{1-x}K_{x}BiO_{3}}$}}
\newcommand{\LACUOD}{{$\mathrm{La_{2}CuO_{4+\delta}}$}}
\newcommand{\YBCO}{{$\mathrm{YBa_{2}Cu_{3}O_{6+\delta}}$}}
\newcommand{\NCCO}{{$\mathrm{Nd_{1.86}Ce_{0.14}CuO_{4+\delta}}$}}
\newcommand{\NCxCO}{{$\mathrm{Nd_{2-x}Ce_{x}CuO_{4+\delta}}$}}
\newcommand{\NCO}{{$\mathrm{Nd_2CuO_{4+\delta}}$}}

\documentclass{ws-p9-75x6-50}

\begin{document}

\title{Electron-phonon interaction in
n-doped cuprates: an Inelastic X-ray Scattering study}

\author{M. d'Astuto}
\address{European Synchrotron Radiation Facility,
BP 220, F-38043 Grenoble Cedex, France}

\author{P. K. Mang}
\address{Department of Applied
Physics, Stanford University, Stanford, California 94305, USA}

\author{P. Giura, A. Shukla, A. Mirone, M. Krisch, F. Sette}
\address{European Synchrotron Radiation Facility,
BP 220, F-38043 Grenoble Cedex, France}

\author{P. Ghigna}
\address{Dipartimento di Chimica Fisica ``M. Rolla'',
Un. Pavia, V.le Taramelli 16, I-27100, Pavia, Italy}

\author{M. Braden}
\address{II. Physikalisches Inst.,
Univ. zu K\"{o}ln, Z\"{u}lpicher Str. 77, 50937 K\"{o}ln, Germany}

\author{M. Greven}
\address{Department of Applied Physics and Stanford Synchrotron Radiation
Laboratory, Stanford, California 94305, USA}


\maketitle

\abstracts{\textit{Inelastic x-ray scattering} (IXS) with very
high (meV) energy resolution has become a valuable spectroscopic tool,
complementing the well established coherent \textit{inelastic
neutron scattering} (INS) technique for phonon dispersion
investigations. In the study of crystalline systems IXS is a
viable alternative to INS, especially in cases where only small 
samples are available. 
Using IXS, we have measured the phonon dispersion of
$\mathrm{Nd_{1.86}Ce_{0.14}CuO_{4+\delta}}$ along the $[\xi,0,0]$
and $[\xi,\xi,0]$ \textit{in-plane} directions. Compared to the
undoped parent compound, the two highest longitudinal optical (LO)
phonon branches are shifted to lower energies because of 
Coulomb-screening effects brought about by the doped charge carriers. 
An additional anomalous softening of the highest branch is observed
around $\mathbf{q}=(0.2,0,0)$. This anomalous softening, akin to
what has been observed in other compounds, provides evidence for a
strong electron-phonon coupling in the electron-doped
high-temperature superconductors. }

\section{Introduction}\label{intro}

Inelastic neutron scattering (INS) measurements of phonon dispersions
in the p-type superconductors 
\LASCUO\cite{pintscB,pingin,mcqueeny,pinbrief},
oxygen-doped \LACUOD\cite{lacuod}, and
\YBCO\cite{pingin,reichardt,petrov} show an anomalous softening of
the highest longitudinal optical (LO) phonon branch with doping,
especially along $\mathbf{q}=[\xi,0,0]$ 
in these p-type cuprate superconductors. 
The atomic displacement
assigned to this branch is the Cu-O bond-stretching 
(BS) mode\cite{pingin,raman}.
The interpretation of such an anomaly invokes a strong 
electron-phonon coupling\cite{pingin,mcqueeny},
but it is still a challenge to provide a quantitative model of 
this coupling
(see Ref.~ \cite{pingin,mcqueeny,pinbrief,fil,wang}). 

Similar anomalies also develop with doping in other perovskite oxides, 
such as \LASMNO\cite{lasmno} and \BKBIO\cite{bkbio}, which are
nearly cubic, and in \LASNIO\cite{lasnio}, which is
iso-structural to \LASCUO.
In the case of \BKBIO, there is evidence that the strong
electron-phonon coupling is responsible for both the anharmonic
behaviour of the high-energy LO modes and the high
superconducting transition temperature (30 K) of this material\cite{bkbio}.
However, a microscopic model capable of explaining the 
electronic and lattice dynamical properties of \BKBIO~ is still
lacking\cite{meregalli}.

The effect of an anomalously strong electron-phonon coupling is
still unclear. At this point, one may not even exclude the
possibility that it may be pair-breaking for the d-wave
superconducting order parameter\cite{anderson}. 
Therefore, it is very important
to analyse the phonon anomalies in as many cuprate families as
possible, and to compare these anomalies with the electronic properties. 
In this context, the different character of electron and hole carriers 
is of central importance:  Cu $3d_{x^2-y^2}$ (O
$2p$) for n(p)-type cuprates\cite{ttu}. This difference leads 
to very different electronic structures\cite{armitage}. 
Since the phonon anomalies
are related to a coupling between charge fluctuations and phonons,
charge carriers with different character may induce quite different
electron-phonon interactions. Therefore, it is imperative to extend
studies of the electron-phonon coupling to the n-type cuprates.

X-ray and neutron scattering are complementary 
experimental tools\cite{settescripta}.
An advantage of IXS over INS is the ability to work with
very small samples due to the very small beam size and high
brilliance obtainable at 3$^{\mathrm{rd}}$ generation synchrotron sources. 
Moreover, the scattering volume for x-rays in the energy range
of 10 to 20 keV is strongly limited by photoelectric absorption
for $Z>4$.
In the case of the cuprates, IXS probes a depth between 10 and
100 $\mu$m, comparable
with the lateral size of the x-ray beam 
obtained on the ID16 (or ID28) beam line at the ESRF. 
This allows experiments with
scattering volumes as small as 10$^{-4}$ mm$^3$. 
Therefore, IXS is
the only technique available for
phonon dispersion
measurements of small samples.
This is relevant for the electron-doped superconductor \NCxCO~ 
in which the oxygen non-stoichiometry $\delta$ may vary throughout
the sample as a consequence of the reduction procedure required to
create the superconducting phase. The small scattering volume
ensures that effects due to any composition gradient are minimised.
A further advantage of the short photoelectric absorption length
is the reduction of multiple scattering processes\cite{settescripta}.

Another important difference between INS and IXS is that the
magnetic cross section is negligible for x-rays, whereas it is
comparable to the nuclear cross section for neutrons. 
Therefore, x-rays are essentially insensitive to magnetic excitations in the
THz regime, which can be helpful for separating magnetic
and lattice contributions. Indeed, in \NCxCO~ there
are very strong crystal field excitations with a neutron cross section
about ten times greater than that for phonons, in particular 
in the range from 4 to 8 THz\cite{lynn}.

Other important differences between IXS and INS are: 
a) The x-ray cross section is
highly coherent. In particular, the entire elastic contribution comes
from $\mathcal{S}(\Delta\bf{Q},\Delta\nu)$, where
$\Delta\bf{Q}$ and $\Delta\nu$ are given by the instrumental
resolution. Therefore only static or {\it quasi-static} structural
and/or chemical disorder contributes to the {\it quasi-elastic}
signal. For perfect, pure crystals no elastic signal is present far
from the Bragg reflections; b) $\Delta\bf{Q}$ and $\Delta\nu$ are
decoupled over a wide range of $(\bf{Q},\nu)$. Moreover, the frequency 
(energy) $\Delta\nu$ is essentially independent of the incident 
frequency (energy) $\nu_i$.
This allows access to high $\bf{Q}$ and/or high frequency transfers 
($\ge$ 20 THz) with good resolution in both $\nu$ and $\bf{Q}$. 
Because of the highly collimated incident beam, 
high $\bf{Q}$-resolution is easily obtained; 
c) the shape of the IXS instrumental energy
resolution is not Gaussian as it is for a neutron triple-axis
spectrometer, but Lorentzian. This strongly enhances the tails of
both the elastic and inelastic signal, and tends to prohibit 
IXS from measuring phonons in the presence of a large elastic
signal such as Bragg peaks or strong diffuse scattering. As a
consequence, inelastic data are usually not available at the zone
centre; d) in the case of high-energy optical branches in the cuprates, 
the relevant scattering is due to oxygen motion, while the 
relevant absorption cross section is due to neodymium. 
As a consequence, the signal from the high-energy optical branches, 
which is relatively low because of the $\frac{1}{\omega}$ terms of the 
dynamical structure factor and of the Bose term, 
is further depressed  compared to the signal of the low-energy modes 
and the elastic signal,
to which all the atoms contribute. 
The strongly enhanced tails of the elastic and 
low-energy inelastic signal (point c)) and Z contrast (point d)) 
above have posed great difficulties for IXS measurements of high-energy
optical branches in transition metal oxides. 
We will show in the next section that we have
been able to overcome these difficulties through a
careful choice of thermodynamic and kinematic conditions.

\section{Experiment and Results}\label{exp}

The inelastic x-ray scattering experiment was carried 
out at the very-high-energy-resolution
beam-line ID16 at the European Synchrotron Radiation Facility (ESRF). 
X-rays from an undulator source were monochromated using a
Si (111) double-crystal monochromator, followed by a
high-energy-resolution backscattering monochromator
\cite{ixsmono} operating at 15816 eV (Si (888) reflection order)
and 17794 eV (Si (999) reflection order). A toroidal gold-coated
mirror refocused the x-ray beam onto the sample, where a beam size
of $250 \times 250~ \mu$m$^2$ full-width-half-maximum (FWHM) was
obtained. The scattered photons were energy-analysed by a
spherical silicon crystal analyser of 3 m radius, operating at the
same Bragg reflection as the monochromator \cite{ixsana}. The
total energy resolution was 1.6 THz (6.6 meV) FWHM for the Si
(888) reflection and 1.1 THz (4.4 meV) FWHM for the Si (999)
reflection. The momentum transfer $\bf{Q}$ was selected by
rotating the 3 m spectrometer arm in the scattering plane
perpendicular to the linear x-ray polarisation vector of the
incident beam. 
For most of the scans, the momentum resolution was set to $\approx
0.087~\mathrm{\AA^{-1}}$ in both the horizontal and the vertical
direction by an aperture of $h \times v = 20 \times 20$ mm$^2$ 
of the slits in front of the analyser. 
In order to achieve higher $\mathbf{Q}$-resolution, for some selected scans  
we set the aperture to  
$20\times10$ mm$^2$ and $10\times10$ mm$^2$. 
Further experimental details are given elsewhere 
(see Ref.~ \cite{ixsmono,ixsana} and references therein).

The samples are two single crystals grown by the travelling-solvent
floating-zone method in 4 atm of $\mathrm{O_2}$ at Stanford
University: a crystal of the parent insulator 
\NCO~ and a crystal of the Ce-doped
superconductor \NCCO. 
The latter sample was oxygen-reduced after the growth and had a
superconducting onset temperature of 
$T_c=24$ K. Further details are given elsewhere\cite{ncco}. 
The samples are of very good crystalline quality. Initial
characterisation work, performed on beam-line 7-2 at the Stanford
Synchrotron Radiation Laboratory (SSRL), gave rocking curve widths 
of 0.02$^{\circ}$ (FWHM) around the $[h,0,0]$
direction.

The IXS experiment was performed in reflection geometry, and the
probed scattering volume corresponded to about $1.5\times10^{-3}$
mm$^3$. IXS scans were performed in the -2$<\nu<$24 THz range, in
the $\mathbf{\tau}=(6,0,0)$, $\mathbf{\tau}=(7,1,0)$, and
$\mathbf{\tau}=(5,5,0)$ Brillouin zones.\footnote{All reciprocal
lattice vectors are expressed in reciprocal lattice units.} 
The samples were mounted on the cold finger of a closed-cycle
helium cryostat, and cooled to 15 K.
The low temperature and high momentum transfer were chosen so as to
optimise the count rate on the high-frequency optical mode while
limiting the loss of contrast due to the contribution from the
tails of the intense low-frequency acoustic modes. 
The effect of lowering the temperature to 15 K and of the  
high momentum transfer is to enhance all the 
signals by the Debye--Waller factor, and to reduce 
the intensities of the low-energy (kT $<h\nu<$ kT$_{RT}$) 
phonons by the Bose factor. 
This is shown in Figs.~ \ref{figa} (a) and (b), 
which compare data obtained at 300 K and 15 K.
We note that the main additional 
contribution at high energy transfers comes 
from the tail of the longitudinal acoustic (LA) 
mode (dashed line), which is strongly reduced at low temperature 
(see  Fig.~ \ref{figa} (b)). 

\begin{figure}[t]
\epsfxsize=30pc 
\epsfbox{satt11fig1_2.eps} 
\caption{IXS phonon spectra. Left panel: for $\mathbf{q}=( 0.1 ,
0.1 , 0 )$ in \NCCO~($\circ$) at (a) 300 K and (b) 15 K and (c)
\NCO~($\bullet$) at 15 K, with a resolution of 
$\Delta \nu = 1.6$ THz. Right panel: (d) at the  reduced
wave-vector $\mathbf{q}=( 0.5 , 0.5 , 0 )$ in \NCO~($\bullet$) at
15 K, with a resolution of $\Delta\nu=$1.6 THz, and at the  reduced
wave-vector $\mathbf{q}=( 0.45 , 0.45 , 0 )$ in \NCCO~($\circ$) at
15 K with a resolution of (e) $\Delta\nu=$1.6 THz 
and (f) $\Delta\nu=$ 1.1 THz. Lines indicate the best fit with
a harmonic oscillator model. The dashed lines in the left panels
indicate the contribution from the tail of the LA mode.
 \label{figa}}
\end{figure}
\begin{figure}[t]
\epsfxsize=15pc 
\epsfbox{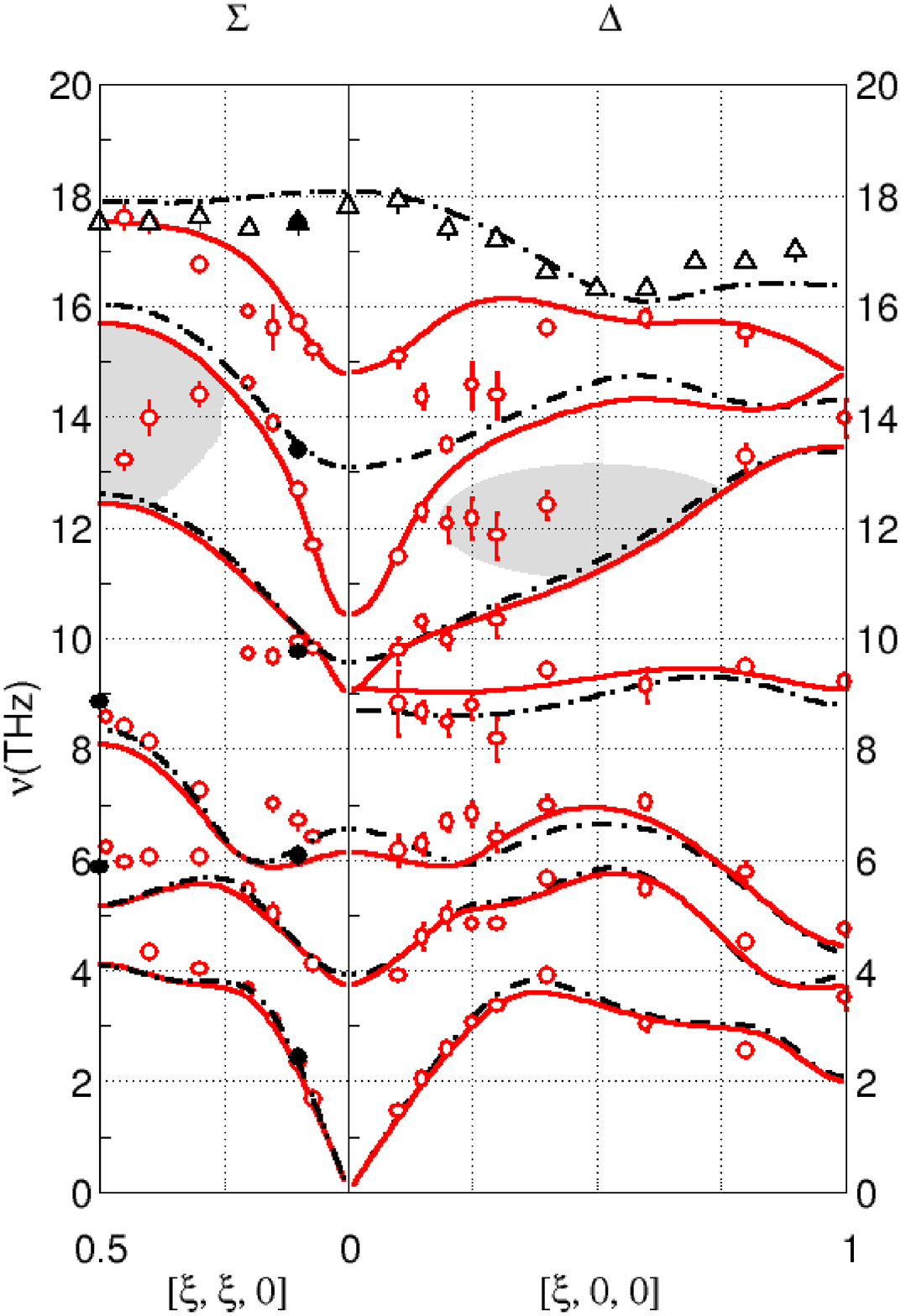} 
\caption{Dispersion of the longitudinal phonon modes in
\NCCO~($\circ$) and \NCO~($\bullet$) measured by IXS. 
For comparison, we include the dispersion of the
highest LO mode in \NCO~($\triangle$) as measured by INS (from Ref.~ 1). 
Solid (dot-dashed) lines
indicate the results of a lattice dynamics calculation 
with a screened (unscreened) Coulomb interaction. 
The shaded areas indicate regions where 
the dynamic structure factor is very
weak and the details of the second highest mode 
can not be determined.\label{figc}}
\end{figure}
The two highest-energy modes for \NCCO~ and \NCO~
at 15 K and at the reduced
wave-vector $\mathbf{q}=( 0.1 , 0.1 , 0 )$
are shown in Fig.~ \ref{figa} (b) and (c), respectively.  
We observe a shift
towards lower energy from the parent antiferromagnetic
insulator to the Ce-doped superconductor. 
The two highest LO branches are assigned to the Cu-O
bond-stretching and O(2) vibration modes\cite{pingin,raman}, 
respectively. While this assignment is correct near the zone
centre, its validity over the whole Brillouin zone is discussed in
Sec.~ \ref{dis}. However, for the sake of simplicity, we refer to
the branches in this way in the text below.

In Fig.~ \ref{figa}, we also show the low-energy acoustic
and optical longitudinal modes at the zone-boundary (M point) in
\NCO~(d) and near the zone-boundary ($\mathbf{q}=( 0.45 ,
0.45 , 0 )$) in \NCCO~(e) and (f). In the doped specimen,
measurements at the M-point were not possible because of the
strong elastic signal at $\tau=(5.5, 5.5, 0)$ 
attributed to a secondary structural phase formed during the reduction
procedure. 
Near the zone boundary, the strongest mode is 
expected to be the lowest-energy longitudinal optical mode. 
We observe the lowest optical mode near the zone-boundary around 6 THz in
both undoped \NCO~and doped \NCCO. This observation is different
from that previously reported for \NCO~based on INS\cite{pintscB}, 
where the same mode was found at lower frequency (about 4 THz). 
We note that in this frequency range, a very strong crystal field
excitation is observed by INS\cite{lynn}. In order to confirm our
assignment we performed a scan with higher energy resolution, 
which is shown in Fig.~ \ref{figa} (f)\footnote{
Note that we observe three acoustic contributions (both longitudinal and
transverse) at low frequency due to the finite {\bf Q}-resolution of our
experiment.}.

The frequencies of the peak positions extracted from these and many
other scans along the main
\textit{in-plane} symmetry directions ($\Delta=(\xi, 0, 0)$ and
$\Sigma=(\xi, \xi, 0)$) are summarised in Fig.~ \ref{figc}. 
The IXS results for \NCCO~ ($\circ$) and \NCO~ ($\bullet$) 
are shown together with the INS data
for the highest Cu-O BS mode ($\triangle$)\cite{pintscB}. 
The lines are the results 
of a numerical calculation\cite{ncco} based on a shell model, 
where the inter-atomic potentials are chosen following 
Ref.~ \cite{chaplot}. We simulated the phonon modes in the parent
antiferromagnetic insulator \NCO~ by adding an unscreened Coulomb
interaction (dot-dashed lines). The results reproduce very
well the experimental dispersion of the undoped parent compound as
measured by INS\cite{pintscB} and IXS.
For the comparison with \NCCO, following Ref.~ \cite{chaplot}, 
we added a screened Coulomb interaction
(solid lines) in order to simulate the effect of the free
carriers introduced by doping\cite{ncco} .

\section{Discussion}\label{dis}

\textit{\textbf{The zone centre. --- }} 
The shift at the zone centre of the high-energy phonon branches of
\NCCO with respect to \NCO is due to the closing of a large 
longitudinal optical - transverse optical (LO-TO)
splitting (also known as \textit{Lyddane--Sachs--Teller} splitting), 
which is very well reproduced by our numerical
calculations. The corresponding $\Delta_1$ and $\Delta_3$
(or $\Sigma_1$ and $\Sigma_3$) branches in \NCO~ are separated by
almost 3 THz at the zone centre, as observed by INS\cite{pintscB},
while for undoped \LACUO \cite{chaplot} the LO and TO branches
are very close. Hence, there does not exist a 
similarly strong screening 
effect in \LASCUO~ for the high-energy \textit{in-plane} LO
modes\cite{pingin}. The screening vector $\kappa_s$, which we found 
to be approximately $0.39~\mathrm{\AA}^{-1}$, is comparable to the one
estimated for \LASCUO \cite{chaplot}.

Although our calculation with a screened Coulomb potential
reproduces the closing of the LO-TO
splitting, we still observe an anomalous additional softening of the
highest LO branch near $\mathbf{q}=(0.2,0,0)$ and
above approximately $\mathbf{q}=(0.2,0.2,0)$
which is not reproduced by our calculations (see Fig.~ \ref{figc}).

\textit{\textbf{$\mathbf{\Delta}$ (or $\mathbf{[\xi, 0, 0]}$) 
line. --- }} Along the $\Delta$ line ($[\xi, 0, 0]$; right side 
of Fig.~ \ref{figc}), the highest LO branch softens in frequency by about
$\nu_1-\nu_2\approx$ 1.5 THz between 
$\mathbf{q}_1=(0.1, 0, 0)$ and $\mathbf{q}_2=(0.2, 0, 0)$, 
which is a shift comparable to the
anomalous shift observed in \LASCUO~at $q$ = 0.25-0.3
\cite{pintscB,mcqueeny,pinbrief}. Therefore, we believe that this
anomalous softening is of the same nature as the one observed in
p-type $\mathrm{La_{1.85}Sr_{0.15}CuO_{4+\delta}}$. 
For $q~>0.2$, the highest $\Delta$ branch recovers rapidly.
This behaviour is most likely due to an anti-crossing with the second
highest branch which is mainly associated with vibrations of the
O(2) atoms in the $\xi$-direction. Therefore, within a standard
anti-crossing framework, one should interpret the highest
longitudinal intensities for $\mathbf{q}=(0.3, 0, 0)$ and above as
being mainly due to O(2) vibration. Within this scenario, the
second-highest branch increases in frequency in the middle of the
zone as in the un-doped compound. Thus, except for the fact that
the LO-TO gap closes, this mode
seems to be insensitive to doping. The LO bond-stretching mode
just above $\mathbf{q}=(0.2, 0, 0)$ is then found at rather low
energies, $\sim$12 THz, with a flat dispersion, as has been
reported for $\mathrm{La_{1.85}Sr_{0.15}CuO_{4+\delta}}$ for
$q>0.25$\cite{mcqueeny} or  $q>0.3$\cite{pinbrief}, but can not be
unambiguously followed to larger q-values. However, the measured
intensities do not behave as expected for a simple anti-crossing
picture (see Ref.~ \cite{ncco}). In an anti-crossing scenario, one expects 
an exchange of character of the two branches, which would imply an
exchange in the intensities. Instead, we observe an intensity that
agrees with our calculation, which does not include any coupling
between the modes, and hence does not allow for an
anti-crossing.  This appears to 
reflect the fact that we still lack 
a realistic model of the phonon anomaly. 
Furthermore, in the (${\bf Q}$,$\omega$) region indicated by the shaded area, 
the dynamic structure factor is very weak and the mode of interest 
is closer to the tails of the 
low-energy modes. Therefore, we can not resolve 
broadening effects as in the hole-doped
case\cite{pinbrief,reichardt}, or even band
splitting\cite{bkbio,lasnio}. 
Consequently, with the present quality of IXS data, 
the position for the second-highest branch can only be defined 
as the centre-of-mass frequency in this (${\bf Q}$,$\omega$) region. 

\textbf{\textit{$\mathbf{\Sigma}$ (or $\mathbf{[\xi, \xi, 0])}$) line. --- }}
Along the $\Sigma$ line (or $[\xi, \xi, 0]$; left side of Fig.~ \ref{figc}),
the softening of the highest LO branch between $\mathbf{q}=(0.1,0.1,0)$ and 
$\mathbf{q}=(0.2,0.2,0)$ is much less than that 
observed at $\mathbf{q}=(0.2,0,0)$. 
However, here we observe a softening of 
the second branch, down to $\approx$ 13 THz for $\xi>0.2$.
In the framework of the anti-crossing picture, this has to be
interpreted as the softening of the Cu-O BS mode,
but the same caution must be taken 
in the analysis of this lower-energy mode 
in the (Q,$\omega$) region 
indicated by the shaded area as in the analysis of the mode along $\Delta$. 
However, we note that: i) due to the different 
symmetry of the orbitals
involved in n- and p-type doping, it has been predicted that the
bond-stretching mode would be particularly soft at the $[\xi,\xi,0]$ zone
boundary\cite{anderson}, 
and ii) the second phase observed in the superconducting
sample (see Sec.~ \ref{exp}), with peaks at 
$\tau=(h+\frac{1}{2}, h+\frac{1}{2}, 0)$, 
may influence the dynamics near the zone boundary. 

\section{Conclusions}\label{conc}

 The present results on the \textit{in-plane} longitudinal
phonon dispersion in \NCCO, when compared to undoped \NCO\cite{pintscB},
show that the anomalous softening of the highest LO phonon mode previously
observed in hole-doped
compounds\cite{pintscB,pingin,mcqueeny,pinbrief,lacuod,reichardt}
is also present in the electron-doped cuprates. 
This observation supports the
hypothesis\cite{pintscB,pingin,mcqueeny,pinbrief,lacuod,reichardt}
of an electron-phonon coupling origin of this feature.

These results furthermore demonstrate that high-energy resolution
inelastic x-ray scattering has developed into an invaluable tool for the
study of the lattice dynamics of complex transition metal oxides, allowing
measurements on small single crystals which can not be studied by 
the traditional method of inelastic neutron scattering. 

\section*{Acknowledgments}
We acknowledge L. Paolasini and G. Monaco for useful discussions
and H. Casalta for precious help during preliminary tests.
The authors are grateful to D. Gambetti, C. Henriquet and R. Verbeni
for technical help, to J.-L. Hodeau for
help in the crystal orientation
and J. -P. Vassalli for crystal cutting, and S.
Larochelle and A. Mehta for assistance with the initial x-ray sample
characterisation. The initial x-ray sample characterization was carried
out at the Stanford Synchrotron Radiation Laboratory, a national user
facility operated by Stanford University on behalf of the U.S. Department
of Energy, Office of Basic Energy Sciences.
P.K.M. and M.G. are supported
by the U.S. Department of Energy under
Contracts No. DE-FG03-99ER45773 and No. DE-AC03-76SF00515, by NSF
CAREER Award No. DMR-9985067.

\end{document}